\documentclass[prb,showpacs,preprintnumbers,amsmath,amssymb,twocolumn,superscriptaddress,longbibliography]{revtex4-1}
\usepackage{graphicx}
\def\be{\begin{equation}}
\def\ee{\end{equation}}
\def\bea{\begin{eqnarray}}
\def\eea{\end{eqnarray}}

\begin{document}

\title{Quantum Fluctuations in the Non-Fermi Liquid System CeCo$_{2}$Ga$_{8}$ Investigated Using $\mu$SR}

\author{A. Bhattacharyya}
\email{amitava.bhattacharyya@rkmvu.ac.in} 
\address{Department of Physics, Ramakrishna Mission Vivekananda Educational and Research Institute, Belur Math, Howrah 711202, West Bengal, India}
\author{D. T. Adroja} 
\email{devashibhai.adroja@stfc.ac.uk}
\affiliation{ISIS Facility, Rutherford Appleton Laboratory, Chilton, Didcot Oxon, OX11 0QX, United Kingdom} 
\affiliation{Highly Correlated Matter Research Group, Physics Department, University of Johannesburg, PO Box 524, Auckland Park 2006, South Africa}
\author{J. S. Lord}
\affiliation{ISIS Facility, Rutherford Appleton Laboratory, Chilton, Didcot Oxon, OX11 0QX, United Kingdom} 
\author{L. Wang}
\address{Beijing National Laboratory for Condensed Matter Physics, Institute of Physics, Chinese Academy of Sciences, Beijing 100190, China}
\address{School of Physical Sciences, University of
Chinese Academy of Sciences, Beijing 100190, China}
\author{Y. Shi}
\address{Beijing National Laboratory for Condensed Matter Physics, Institute of Physics, Chinese Academy of Sciences, Beijing 100190, China}
\address{School of Physical Sciences, University of
Chinese Academy of Sciences, Beijing 100190, China}
\address{Songshan Lake Materials Laboratory, Dongguan, Guangdong 523808, China}
\author{K. Panda}
\address{Department of Physics, Ramakrishna Mission Vivekananda Educational and Research Institute, Belur Math, Howrah 711202, West Bengal, India}
\author{H. Luo} 
\address{Beijing National Laboratory for Condensed Matter Physics, Institute of Physics, Chinese Academy of Sciences, Beijing 100190, China}
\address{Songshan Lake Materials Laboratory, Dongguan, Guangdong 523808, China}
\address{School of Physical Sciences, University of
Chinese Academy of Sciences, Beijing 100190, China}
\author{A. M. Strydom} 
\address{Highly Correlated Matter Research Group, Physics Department, University of Johannesburg, PO Box 524, Auckland Park 2006, South Africa}

\begin{abstract}

\noindent Reduced dimensionality offers a crucial information in deciding the type of the quantum ground state in heavy fermion materials. Here we have examined stoichiometric CeCo$_{2}$Ga$_{8}$ compound, which crystallizes in a quasi-one-dimensional crystal structure with Ga-Ce-Co chains along the $c$-axis. The low-temperature behavior of magnetic susceptibility ($\chi\sim-\ln T$), heat capacity ($C_p/T\sim-\ln T$), and resistivity ($\rho\sim T^{n}$) firmly confirm the non-Fermi liquid ground state of CeCo$_{2}$Ga$_{8}$. We studied the low-energy spin dynamics of CeCo$_{2}$Ga$_{8}$ compound utilizing zero field (ZF-) and longitudinal field (LF-) muon spin relaxation ($\mu$SR) measurements. ZF-$\mu$SR measurement reveals the absence of long-range magnetic ordering down to 70 mK, and interestingly below 1 K, the electronic relaxation rate sharply rises, intimating the appearance of low energy quantum spin fluctuations in CeCo$_{2}$Ga$_{8}$.  

\end{abstract}

\date{\today} 

\maketitle

\section{Introduction}

\noindent Searching for quantum critical point (QCP) is a great challenge in strongly correlated materials since it only emerges at zero temperature by varying a control parameter such as magnetic field, pressure, and chemical doping or alloying etc~\cite{Lohneysen,Shibauchi2014,Fischer2005}. Heavy fermion (HF) materials exhibit many exotic states in the vicinity of a magnetic QCP, including non-Fermi- liquid (NFL) and unconventional superconductivity~\cite{Lohneysen,Georges,Varma,Coleman,Riseborough,Stewart2001}. At QCP, the quantum fluctuations dominate over the thermal fluctuations that break the predictions of well-known Landau-Fermi-liquid behavior~\cite{Landau1,Landau2}, and hence the system exhibits NFL behavior. The nature of quantum fluctuations and development of magnetic correlations  will depend on the dimensionality of the  systems, and hence it is very important to investigate effect of dimensionality on the QCP/NFL. So far, most Ce-based QCP/NFL systems investigated are two-dimensional (2D) or 3D, and there are no reports on 1D Ce-based NFL systems~\cite{Stewart2001,Krellner,Steppke2013}. Furthermore, the physical properties of Ce-based compounds at low temperature exhibit Fermi liquid behavior predicted by Landau theory~\cite{Landau}. For example, at low-temperature electrical resistivity $\rho \sim T^{2}$, heat capacity $C \sim T$ and dc magnetic susceptibility independent of temperature~\cite{Schofield1999,Stewart2006,Gegenwart2008,Si2010,Stockert}. Interestingly some of the Ce- and Yb- based materials deviate from conventional Fermi liquid behavior to so call NFL behavior, which can be tuned from an antiferromagnetic ground state to zero-temperature QCP, where quantum fluctuations are responsible for NFL behavior. In case of NFL compounds $\rho$ $\sim T^{n}$ (1 $\leq$ n $<$ 2),~$C/T$ $\sim$ $-\ln T$ $C/T$ $\sim$ $a - bT^{1/2}$ and $\chi$ $\sim$ $-\ln T$ or $\chi$ $\sim$ $T^{-p}$ $(p < 1)$~\cite{Schofield1999,Stewart2006,Gegenwart2008,Si2010,Stockert}. 

\noindent To accommodate deeper insight toward the specific nature of the QCP, both theoretical and experimental efforts have been made a lot in recent years. Still, most of them focus on the quasi-2D or 3D HF~\cite{Stewart2001,Krellner,Steppke2013}. Considering dimensionality is a fundamental component in defining the unique NFL attributes in these materials, lower dimension anticipates a substantial magnetic frustration parameter, and it is imperative to search QCP in quasi-1D HF compounds, whose science can be further easier approximated by the density matrix renormalization group method as well as mean-field approximation~\cite{Schol}. QCP has been observed in CeCu$_{6-x}$T$_{x}$ (T = Au, Ag)~\cite{Lohn,Heuser}, in which the heavy NFL at x$_{c}$ = 0.1 at an ambient pressure is driven to a magnetically ordered state via further doping, in antiferromagnetic ordered HF compounds such as CeIn$_{3}$~\cite{Shishido,Gor} or CePd$_{2}$Si$_{2}$ ~\cite{Mathur}. To date, there exist hardly a few undoped or stoichiometric materials which exhibit NFL states at ambient pressure such as UBe$_{13}$~\cite{Ramirez,Bommeli}, CeNi$_{2}$Ge$_{2}$~\cite{Steglich1997} and CeCu$_{2}$Si$_{2}$~\cite{Steglich1997}, CeRhBi~\cite{Sasa,Anand}. 

\begin{figure*}[t]
\centering
 \includegraphics[height=0.6\linewidth,width=0.9\linewidth]{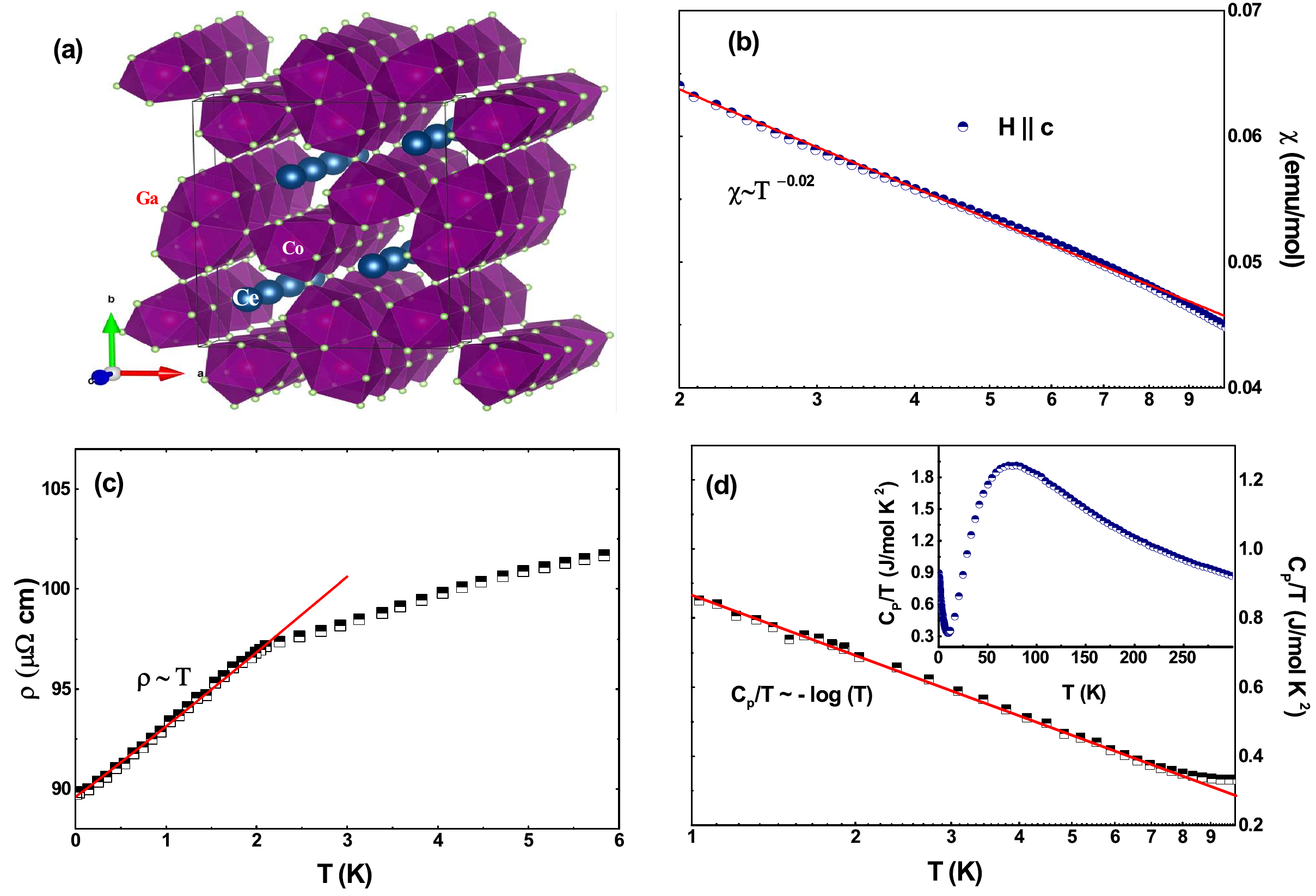}\hfil
\caption {(a) Represent the crystal structure of CeCo$_{2}$Ga$_{8}$, showing the quasi-1D chains of Ce atoms along the $c$-axis. The individual unit cell holds four Ce atoms. (b) Magnetic susceptibility as a function of temperature along the parallel to $c$-axis ($\chi \sim -\ln{T}$). The solid line shows the power law fit. (c) Low temperature region of resistivity ($\rho \sim T$). (e) Low temperature dependence heat capacity divided by temperature $C_{p}/T$ plotted in semi-logarithmic scale in zero applied field ($C_{p}/T \sim -\ln{T}$). Inset displays $C_{p}/T$ as a function over a large temperature range.}
\label{physical properties}
\end{figure*}

\noindent As chemically disorder NFL states remain challenging to understand by theoretical models, it is highly desirable to examine stoichiometric and homogeneous systems, which are prototypical materials for theoretical modeling. The recently discovered quasi-1D Kondo lattice system CeCo$_{2}$Ga$_{8}$, which crystallizes in the YbCo$_{2}$Ga$_{8}$-type orthorhombic structure, provides us a rare opportunity to scrutinize a QCP at ambient pressure~\cite{Wang}. The onset of coherence is about $T^* \sim$ 20 K, and no sign of superconductivity is found down to 0.1 K. Furthermore, 1D spin-chain behavior is also clear from susceptibility data~\cite{Wang}, and density functional computations predict flat Fermi surfaces originating from the 1D $f$-electron bands along the $c$-axis. NFL state develops in a wide temperature region, as apparent from the pressure dependence resistivity data~\cite{Wang}. All these facts firmly intimate CeCo$_{2}$Ga$_{8}$ is naturally positioned in the proximity of a magnetic QCP~\cite{Wang}. Nevertheless, the $T$-linear resistivity and a logarithmically divergent-specific heat is expected in the 2D antiferromagnetic QCP from the conventional Hertz-Millis theory~\cite{Hertz1976,Millis1993}, but not in a quasi-1D Kondo lattice system. Interestingly in the case of CeCo$_{2}$Ga$_{8}$, anisotropic magnetic attributes well explained utilizing crystal field theory, and the ratio of the exchange interaction is $|J_{ex}^c/J_{ex}^{a,b}|\sim$ 4-5~\cite{Cheng}. Our investigation on CeCo$_{2}$Ga$_{8}$ includes electrical resistivity $\rho$(T), dc susceptibility $\chi(T)$, heat capacity $C_\mathrm{P}$(T) data used for characterization of CeCo$_{2}$Ga$_{8}$, and muon spin relaxation ($\mu$SR) measurement to study the low energy spin dynamics. Our microscopic examination confirms the stoichiometric CeCo$_{2}$Ga$_{8}$ compound exhibits NFL ground state without any doping. It is interesting to note that NFL ground state without doping is rare in Ce-based compounds, only a few CeRhBi~\cite{Sasa,Anand}, CeRhSn~\cite{Ho}, CeInPt$_{4}$~\cite{Malik,Hillier2007} located in this group.

\section{Experimental Details}

\noindent For the present investigation, a high-quality single crystals of CeCo$_{2}$Ga$_{8}$ were grown employing the Ga-flux method. The complete growth method can be found in Ref.~\cite{Wang}. Temperature-dependent magnetic susceptibility [$\chi(T)$] was measured using a Quantum Design Magnetic Property Measurement System (MPMS-SQUID) with the applied field parallel to the $c$-axis. Electrical resistivity [$\rho(T)$] and heat capacity [$C_p(T)$] measurements were done using a Physical Property Measurement System (PPMS) with He$^{3}$ cryostat. 

\par

\begin{figure*}[t]
\centering
 \includegraphics[height=0.6\linewidth,width=0.9\linewidth]{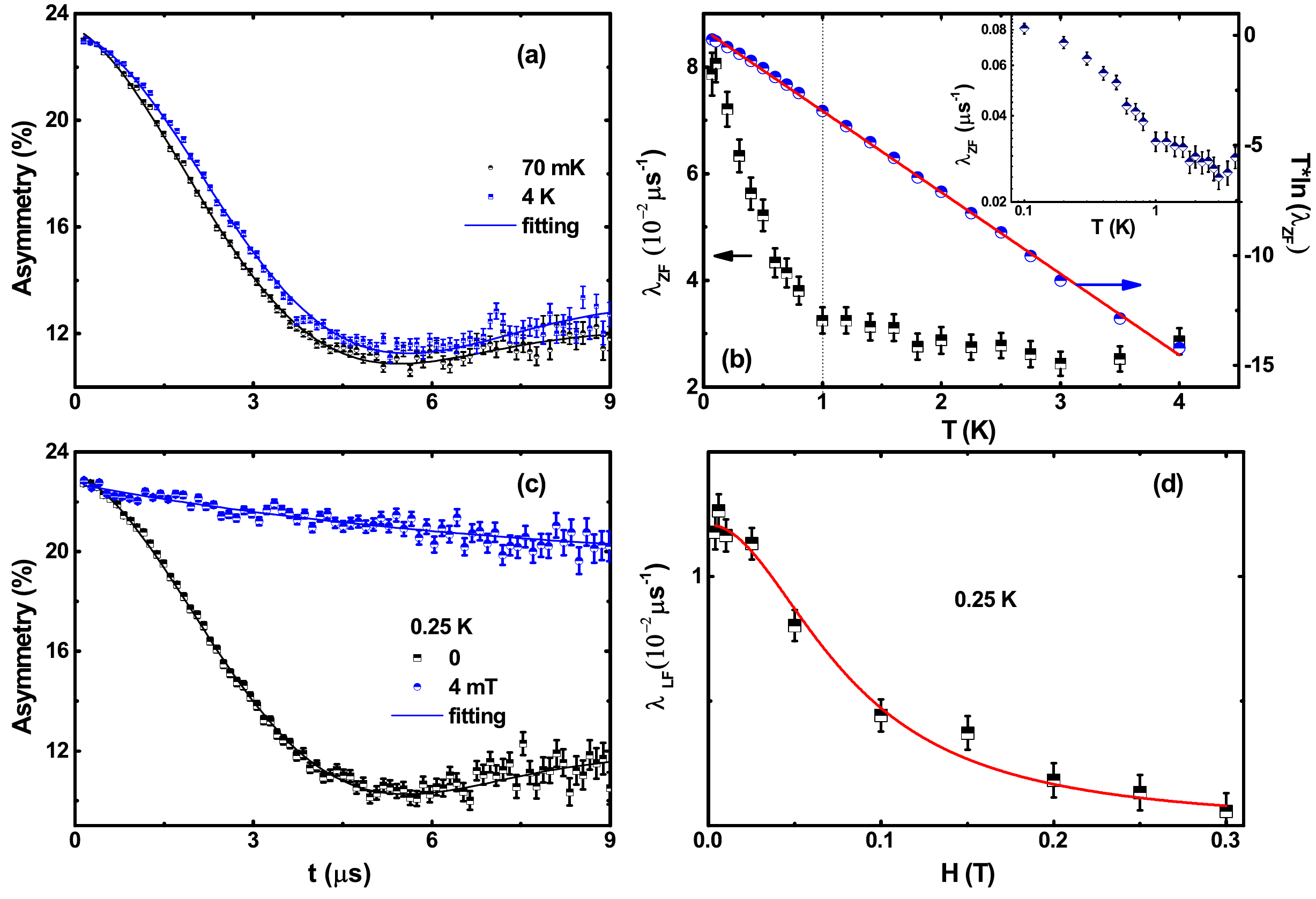}\hfil
\caption {(a) Time dependence zero-field $\mu$SR spectra of CeCo$_{2}$Ga$_{8}$ collected at 70 mK (black), and 4 K (blue). Solid lines are the least-squares fit applying Eq. 1. (b) The left y-axis outlines the electronic contribution of the muon relaxation rate $\lambda_{ZF}$. A clear signature of quantum fluctuations is seen below 1 K, confirm the NFL state as marked in the bulk properties. The right panel demonstrates the Arrhenius behavior of $\lambda$. The line is the least-square fit of the data, as presented in the text. Inset shows the log-log plot of $\lambda_{ZF}$ with temperature. (c) Time dependence longitudinal field muon asymmetry spectra of CeCo$_{2}$Ga$_{8}$ measured at $T$ = 0.25 K in zero field and 40 mT. Solid red lines are the least-square fit to the raw data using Eq. 1. (d) Describes the longitudinal field dependence of the muon relaxation rate $\lambda$ for CeCo$_{2}$Ga$_{8}$ at 0.25 K. The solid red line is the least-square fit using Eq. 2.}
\label{musr}
\end{figure*}

\noindent $\mu$SR experiments were carried out using the MUSR spectrometer in the zero-field (ZF) and longitudinal field (LF, $H~\parallel$ muon spin direction) geometries at the muon beamline of the ISIS Facility at the Rutherford Appleton Laboratory, United Kingdom~\cite{doi}. The unaligned single crystals [as they were very small in size less than 1.5mm (length) $\times$ 1mm (diameter)] of CeCo$_{2}$Ga$_{8}$ were mounted on a 99.995\% silver plate applying thinned GE varnish covered with a high purity silver foil. The sample was cooled down to 70 mK using a dilution refrigerator. Spin-polarized incident muon thermalizes on the sample, and the resultant asymmetry determined employing, $G_{\mathrm{z}}\left(t\right) = [{N_{\mathrm{F}}\left(t\right)-\alpha N_{\mathrm{B}}\left(t\right)}]/[{N_{\mathrm{F}}\left(t\right)+\alpha N_{\mathrm{B}}\left(t\right)}]$, where $N_{\mathrm{B}}\left(t\right)$ and $N_{\mathrm{F}}\left(t\right)$ are the number of positrons counted in the backward and forward detectors respectively and $\alpha$ is an instrumental calibration constant determined with a small (2~mT) transverse magnetic field at high temperature. $G_{z}$(t) gives information about the spin-lattice relaxation rate and internal field distribution. We have collected ZF $\mu$SR data between 0.07 K and 4 K and LF $\mu$SR data in an applied field up to 0.3 T at 0.25 K. ZF/LF-$\mu$SR data were analyzed utilizing WiMDA software~\cite{Pratt2000}.

\section{Results and discussion}

\noindent Figs.~\ref{physical properties}(a) represent the orthorhombic structure (space-group $Pbam$, No. 55) of CeCo$_{2}$Ga$_{8}$, showing the quasi-1D chains of Ce atoms along the $c$-axis. The individual unit cell holds four Ce atoms. Non-Fermi liquid state in CeCo$_{2}$Ga$_{8}$ is confirmed by electrical resistivity, magnetization, and heat capacity measurements, and the results are presented in Figs.~\ref{physical properties}(b-d). The temperature variation of susceptibility measured in a field cool condition with 50 mT applied field parallel to the $c-$ axis is manifested in Fig.~\ref{physical properties}(b). $\chi(T)$ manifests a power-law response suggesting CeCo$_{2}$Ga$_{8}$ placed near the quantum phase transition~\cite{Wang}. High temperature Curie-Weiss fit yields an effective magnetic moment $p_{eff}$ = 2.74 $\mu_{B}$, which is larger compare to free Ce$^{3+}$-ion value (2.54 $\mu_{B}$), may indicate very  weak magnetic contribution from the Co ion in CeCo$_{2}$Ga$_{8}$. This peculiarity is well-known in Ce-based heavy Fermion systems, for instance, CeCoAsO~\cite{Rajib} and CeCo$_2$As$_2$~\cite{Stevens}. In low temperature limit as shown in Fig.~\ref{physical properties}(b), 0.1 K $\leq T \leq$ 2 K, $\rho(T)$ varies linearly with temperature ($\rho \sim$ $T$), characteristics of a NFL state. Heat capacity varies logarithmically $C_{p}/T$ $\sim-\ln(T)$ in the low $T$ limit as shown in Fig.~\ref{physical properties}(d). It is clear from the inset of \ref{physical properties}(d) that the heat capacity exhibits a broad maximum near 90 K, which can the attributed to the Schottky anomaly arising from the crystal field effect in the presence of Kondo effect. All of these attributes of CeCo$_{2}$Ga$_{8}$ are pretty similar to quasi 1D NFL ground state and lead to further examination using a microscopic technique such as muon spin relaxation ($\mu$SR) measurement.

\par

\noindent To probe the NFL state as seen from electrical, thermal and magnetic measurements at low $T$, we did ZF/LF-$\mu$SR measurements. The ZF depolarization reveals the sum of the local responses of muons embedded at different stopping sites in CeCo$_{2}$Ga$_{8}$. ZF-$\mu$SR muon asymmetry spectra of CeCo$_{2}$Ga$_{8}$ at $T$ = 70 mK (black) and $T$ = 4 K (blue), that are representative of the data collected, are shown in Fig.~\ref{musr}(a). Both the 4 K and 70 mK data in Fig.~\ref{musr}(a) reveal the same value of the initial asymmetry at t=0 along with the lack of oscillations confirm the absence of long range magnetic ordering in CeCo$_{2}$Ga$_{8}$ down to 70 mK. This precludes any argument for static magnetism in the sample~\cite{Suzuki}. Hence the moderate increase of the relaxation upon cooling from high temperature reflects only a slowing down of the electronic spin dynamics. Fits to the ZF-$\mu$SR data at different temperatures were done employing a Gaussian Kubo-Toyabe function multiplied by an exponential decaying function~\cite{Bhattacharyya}, 

\begin{equation}
P_{z}(t)=A_{1}[\frac{1}{3}+\frac{2}{3}(1-\sigma^{2}_{KT}t^{2})\exp^{(\frac{-\sigma^{2}_{KT}t^{2}}{2}}]\exp{(-\lambda_{ZF} t)} + A_{bg}
\label{ZF}
\end{equation} 

\noindent where $\lambda_{ZF}$ is the ZF relaxation rate arising due to the local moment, $A_{1}$ and $A_{bg}$ are the asymmetries were originating from sample and background, respectively. $A_{bg}$ was determined from high-temperature ZF, which was kept fixed for the analysis. $\sigma_{KT}$ is the nuclear contribution that emerges from the Gaussian distribution of the magnetic field at the muon site. The relaxation term $\exp$(-$\lambda_{ZF} t$) is the magnetic contribution that comes from the fluctuating electronic spins, which provides information about the low energy spin dynamics of CeCo$_{2}$Ga$_{8}$. As shown in the left panel of Fig.~\ref{musr}(b), the $T$ variation of $\lambda_{ZF}$ sharply increases below 1 K, indicating the development of NFL state as evidence from bulk properties. Above 1 K, $\lambda_{ZF}$ decreases with increasing temperature. The right panel of Fig.~\ref{musr}(b) represents the Arrhenius like behavior of $\lambda_{ZF}(T)$, i.e., follows the form, $\lambda_{ZF} = \lambda_{0}\exp (-\frac{E_{a}}{k_BT})$, where $E_{a}$ and $k_B$ are the activation energy and Boltzmann constant respectively. This confirms that the low $T$ spin dynamics of CeCo$_{2}$Ga$_{8}$ is a thermally activated with $E_{a}$ = 2.3 mK which is similar as observed for CeInPt$_{4}$~\cite{Hillier2007} and CeRhBi~\cite{Anand} with $E_{a}$ values are 2.9 mK and 140 mK, respectively. It is an open question why the temperature dependence relaxation of these three stoichiometry compounds exhibits Arrhenius behavior in NFL state (as T$\to$ 0), while that of chemically disorder NFL systems it exhibits power law behavior~\cite{Adroja2008}. We also plotted $\lambda_\mathrm{ZF}$(T) data of CeCo$_{2}$Ga$_{8}$ in log-log plot [Inset of Fig.~\ref{musr} (b)] to see power law behavior, but the data did not follow the power law-behavior. 

\par

\noindent A longitudinal field of just 40 mT removes any relaxation due to the spontaneous field and is adequate to decouple muons from the relaxation channel, as presented in Fig.~\ref{musr}(c) at 250 mK. Once muon is decoupled from the nuclear moments, the spectra can be best-fitted using~\cite{Kubo}, $G_{z(t)}=A_{1}\exp{(-\lambda_{LF}t)}+A_{bg}$. $\lambda_{LF}$ decreases rapidly at low $H$ and saturates at a high $H$. $\lambda_{LF}(H)$ can be adequately expressed by the standard description description given by the Redfield formula~\cite{Alian},

\begin{equation}
\lambda = \lambda_0+\frac{2\gamma^2_{\mu}<H_{l}^2>\tau_C}{1+\gamma^2_{\mu}H^2\tau_C}
\end{equation}

\noindent where $\lambda_0$ is the field independent depolarization rate, $<H_{l}^2>$ is the time-varying local field at muon sites due to the fluctuations of Ce 4$f$ moments. $H_{l}$ is applied longitudinal field and  the correlation time $\tau_\mathrm{C}$ is related to the imaginary component of the $q$-independent dynamical susceptibility, $\chi''(w)$ through the fluctuation-dissipation theorem~\cite{Toll}. The red line in Fig.~\ref{musr}(d) represents the fit to the $\lambda_{LF}(H)$ data. The calculated parameters are $\lambda_0$ = 0.19(1) $\mu s^{-1}$, $<H_{l}^2>$ = 1.3(1) mT, and $\tau_c$ = 3.1(6)$\times$10$^{-8}$ s. The value of the time constant of CeCo$_{2}$Ga$_{8}$ unveils a slow spin dynamics, which is originated by the quantum critical fluctuations at low $T$. A similar values are observed for CeRhBi~\cite{Anand}, $\lambda_0$ = 0.17(1) $\mu s^{-1}$, $<H_{l}^2>$ = 1.5(1) mT, and $\tau_c$ = 4.2(6)$\times$10$^{-8}$ s. 

\section{Conclusion}

\noindent In conclusion, we have presented the magnetization, resistivity, heat capacity, and ZF/LF muon spin relaxation measurements on the quasi one-dimensional CeCo$_{2}$Ga$_{8}$ compound. The linear behavior of $\rho(T)$ and logarithmic divergence of $C_p(T)/T$ imply a NFL ground state of CeCo$_{2}$Ga$_{8}$. Moreover, the increase in the ZF relaxation rate $\lambda_{ZF}$ below 1 K and the Arrhenius like behavior of CeCo$_{2}$Ga$_{8}$ suggest NFL ground state, which is quite similar to that seen in other NFL systems, CeRhBi~\cite{Anand} and CeInPt$_{4}$~\cite{Hillier2007}. Our ZF $\mu$SR measurements  confirm the absence of long range magnetic ordering down to 70 mK. Furthermore, the longitudinal field dependence $\mu$SR study provides information on the spin fluctuations rate and the width of field distribution at the muon sites. The observed quantum fluctuations below 1 K in the undoped CeCo$_{2}$Ga$_{8}$ compound makes it a prototype material to investigate the low $T$ quantum fluctuations in low dimensional NFL systems and other Ce128 counterparts with YbCo$_{2}$Ga$_{8}$-type structure. This work will pave the way in our understanding of NFL in 1D systems both theoretically and experimentally.

\section{Acknowledgments}

\noindent We would like to thank Dr. J. Sannigrahi and Dr. V. K. Anand for interesting discussions. A. B. would like to thank DST India for Inspire Faculty Research Grant (DST/INSPIRE/04/2015/000169). D. T. A. would like to acknowledge CMPC-STFC, grant number CMPC-09108. D. T. A. gratefulness to JSPS for invitation fellowship and the Royal Society of London for UK-China Newton mobility grant. K. P. would like to acknowledge DST India, for an Inspire Fellowship (IF170620). A. B. thank the Department of Science and Technology, India (SR/NM/Z-07/2015) for the financial support and Jawaharlal Nehru Centre for Advanced Scientific Research (JNCASR) for managing the project. Work at IOP, CAS is supported by  by the National Key Research and Development Program of China (2017YFA0303100, 2017YFA0302900), the National Natural Science Foundation of China (11974392, 11822411, 11774399 and 11961160699),  and Beijing Natural Science Foundation (JQ19002, Z180008). H. L. is grateful for the support from the Youth Innovation Promotion Association of CAS (2016004). We would like to thank the ISIS Facility for beam time, RB1810451~\cite{RB}.

\end{document}